\documentclass[final,1p,times,twocolumn]{elsarticle}

\usepackage{graphicx}
\usepackage{epsfig}

\usepackage{amssymb}

\journal{Physica A}

\begin{document}

\begin{frontmatter}

\title{Describing the ground state of quantum systems through statistical mechanics}
\author{Andre M. C. Souza}

\address{Departamento de Fisica, Universidade Federal de
Sergipe, 49100-000 Sao Cristovao-SE, Brazil \\
National Institute of Science and Technology for Complex Systems,
Rua Xavier Sigaud 150, 22290-180 Rio de Janeiro-RJ, Brazil}

\ead{amcsouza@ufs.br}

\begin{abstract}
We present a statistical mechanics description to study the ground
state of quantum systems. In this approach, averages for the
complete system are calculated over the non-interacting energy
levels. Taking different interaction parameter, the particles of the
system fall into non-interacting microstates, corresponding to
different occupation probabilities for these energy levels. Using
this novel thermodynamic interpretation we study the Hubbard model
for the case of two electrons in two sites and for the half-filled
band on a one-dimensional lattice. We show that the form of the
entropy depends on the specific system considered.

\end{abstract}


\end{frontmatter}


\section{Introduction}
\label{}

Statistical Mechanics (SM) provides useful concepts to study systems
with large number of particles. For example, based on standard SM,
Edwards \cite{edw} proposed a thermodynamic description of granular
matter in which thermodynamic quantities are computed as flat
averages over configurations where the grains are static or jammed,
leading to a definition of configurational temperature. A numerical
diffusion-mobility experiment of a granular system has supported the
Edwards' statistical ensemble idea \cite{makse}. Another example is
the relation between entropy and the horizon area of a black hole
\cite{black}, which provides a new approach for studying black holes
and quantum gravity theory. Furthermore, four laws of black hole
mechanics can be demonstrated using this thermodynamic description.
The microscopic origin of the black hole entropy, originally
calculated thermodynamically, has been explained from string theory.
\cite{black2}

Recently, Cejnar et al. \cite{cej} analyzed quantum phase
transitions in finite systems \cite{bor1} by defining an analog of
the absolute temperature scale connected to the interaction
parameter of the Hamiltonian. And thus, they were capable of
establishing a thermodynamic analogy for the quantum phase
transition. However, they did not identify the correspondence with
statistical mechanics and consequently the new scenario opened by
this microscopic analysis. This correspondence and these
consequences are the goal of this paper.

Here, we use tools developed in SM to study the ground-state of
quantum systems. We observe that, for certain classes of quantum
systems, taking different intensities of the interaction between
particles of the system corresponds to taking different occupation
probabilities for non-interacting microstates energy levels. With
this observation we can define an analog of the absolute temperature
scale in such a manner that it is possible to make a thermodynamic
interpretation for the interaction in the ground-state of quantum
systems. The Hubbard Hamiltonian is a typical model in which this
approach can be applied. Here, we analyze two exact solvable limits
of the Hubbard model.

This paper is organized as follow. The formalism is described in
Sec. \ref{form}. The study of the two exact solvable problems based
on the Hubbard model are presented in Sec. \ref{app}. Finally, we
present the conclusions in Sec. \ref{concl}.

\section{Formalism}
\label{form}

The scheme of our formalism can be applied to a broad class of
Hamiltonians defined as
\begin{equation} \label{e1}
\hat{H}=\hat{H}_{0} + T \hat{V},
\end{equation}
where we assume that $\hat{H}_{0}$ is a one-particle Hamiltonian
operator and the interaction term is given by the $\hat{V}$ operator
and $T$ is the dimensionless interaction parameter. Here, we must
consider that $T \ge 0$ and the operator $\hat{V}$ is positively
defined. In this way we have established that the energy is a
concave function of $T$

A good example of this class of Hamiltonians is the one of the
Hubbard model \cite{hub}. In this model, which is amongst the most
important magnetic ones, the eigenstates of the Hamiltonian in the
absence of interaction ($T=0$) are just the non-interacting states
$| \phi_{i} \rangle $, whose respective energy
 eigenvalues, $E_i (0)$ are defined through the relation $\hat{H}_{0} | \phi_{i} \rangle
=E_i (0) | \phi_{i} \rangle$. The eigenvalues $E_i (T)$ of $\hat{H}$
for nonvanishing $T$ are obtained from the equation $\hat{H} |
\psi_{i} \rangle = E_i (T) | \psi_{i} \rangle$. Moreover, the
expectation value of an operator $\hat{X}$ on the ground-state $|
\psi_{0} \rangle$ is given by $\langle \psi_{0} | \hat{X} | \psi_{0}
\rangle$.

Now, we can provide a approach for obtaining expectation values of
physical quantities on the ground-state in the base of
non-interacting states. This simply means to find the expectation
values of $\hat{X}$ on the ground-state in the $| \phi_{i} \rangle $
representation.

The ground-state $| \psi_{0} \rangle$ can be expanded in terms of
the non-interacting states $| \phi_{i} \rangle $ as
\begin{equation}
| \psi_{0} \rangle = \sum_{i} a_i (T) | \phi_{i} \rangle,
\end{equation}
where the coefficients $a_i (T) = \langle  \phi_{i} | \psi_{0}
\rangle$. We recall that the quantity $|a_{i} (T) |^{2}$ has a
{\em{probabilistic}} interpretation. In other words, we can write
$p_{i} (T) \equiv |a_{i} (T) |^{2} \; \epsilon \; [0,1]$ and
$\sum_{i} p_{i} (T) = \sum_{i} |a_{i} (T) |^{2} = 1$. This
establishes the connection to SM: the expectation value of
$\hat{H}_{0}$,
\begin{equation}
\langle \hat{H}_{0} (T) \rangle = \langle \psi_{0} | \hat{H}_{0} |
\psi_{0} \rangle = \sum_{i} p_i (T) E_i (0),
\end{equation}
can be interpreted as a usual average, which is computed over the
set of non-interacting energy levels $E_{i}(0)$, each one
 with probability $p_i (T)$. It is an analog of the mean
energy $ \langle E \rangle =\sum_{i}{p_i E_i}$. We can easily verify
that if $T \ge 0$ and $\langle \hat{V} (T) \rangle \ge 0$ is a
monotonically decreasing function of $T$, then $E_{0} (T_{1}) \le
E_{0} (T_{2})$ for $T_{1} \le T_{2} $. In this case, in analogy to
SM, for the non-interacting case $T=0$, the system has the lowest
energy $E_{0} (0)$ and $p_i (0)= \delta_{i0}$. If $T >0$, like a
thermal energy, the interaction favors other energy levels of the
non-interacting case. In this description, only the non-interacting
microscopic states are used to compute the {\em thermodynamic}
properties. This enables us to define an analog of the absolute
temperature scale, called ground-state temperature, as $T_{g}=T/k$,
where $k$ is a constant measured in Kelvins$^{-1}$.

\begin{figure}
\includegraphics[width=85mm]{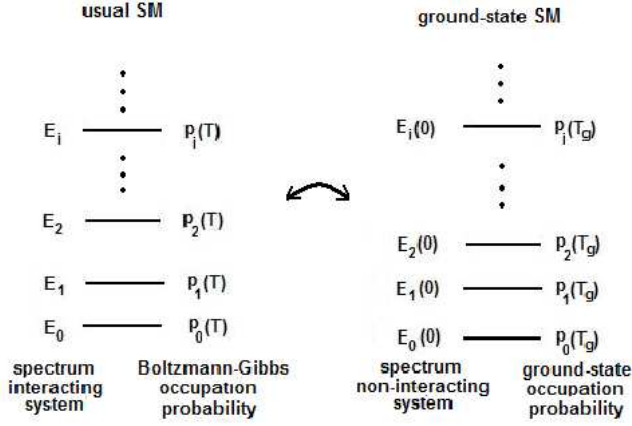}
\caption{Similar to the usual canonical ensemble of the SM, taking
different ground-state temperature $T_{g}$ (interaction parameter),
the particles of the system fall into non-interacting microstates,
corresponding to different occupation probabilities for these energy
levels.} \label{fi1}
\end{figure}

This description is illustrated in Fig. \ref{fi1}. Similar to the
usual canonical ensemble of the SM, we can consider that taking
different ground-state temperatures $T_{g}$, i.e, different values
of the interaction parameter, the particles of the system fall into
non-interacting microstates, corresponding to different occupation
probabilities for these energy levels.

In addition, an analogy to the standard thermodynamics is also
reproduced by this description. We can introduce a so-called
ground-state thermodynamics, defining the ground-state internal
energy, ground-state free energy and ground-state entropy,
respectively, as
\begin{equation} \label{ug}
U (T_{g}) = \langle E (T_{g}) \rangle = \sum_{i} p_i (T_{g}) E_i
(0),
\end{equation}
\begin{equation} \label{fg}
F (T_{g}) = E_{0} (T_{g}) - T_{g} \langle \hat{V} (0) \rangle,
\end{equation}
\begin{equation} \label{sg}
S (T_{g}) = k ( \langle \hat{V} (0) \rangle - \langle \hat{V}
(T_{g}) \rangle ).
\end{equation}

It can be easily seen that $S (T_{g})$ is a non-negative
monotonically increasing function of $T_{g}$. We can trivially
verify that, using Eqs. (\ref{ug})-(\ref{sg}), the ground-state
thermodynamics precisely satisfies the standard thermodynamics
relation for the Helmholtz free energy
\begin{equation} \label{futs}
F (T_{g}) = U (T_{g}) - T_{g} \; S(T_{g}).
\end{equation}
Furthermore, we can derive the thermal response function, in
correspondence to the heat capacity
\begin{equation} \label{sh}
C (T_{g}) = T_{g} \frac{dS(T_{g})}{dT_{g}} = -T_{g}
\frac{d^{2}F(T_{g})}{dT_{g}^{2}}.
\end{equation}
It is interesting to observe that the expression above can be
calculated using the Hellmann-Feynman theorem which allows to find
the ground-state expectation values of a general operator $\hat{X}$
by differentiating the ground state energy of a perturbed
Hamiltonian $\hat{H}_{0}+\lambda \hat{X}$ with respect to $\lambda$
\cite{sor}.

\section{Applications}
\label{app}

For illustrating the approach introduced in this letter, let us
study two exact solvable problems based on the Hubbard model
\cite{hub}. The Hamiltonian of the Hubbard model is defined as
\begin{equation}
\hat{H}= -t \sum_{\langle ij \rangle \alpha}
\hat{c}_{i\alpha}^{\dag} \hat{c}_{j\alpha} + U \sum_{i}
\hat{n}_{i\uparrow} \hat{n}_{i\downarrow} , \label{hamil}
\end{equation}
where $\hat{c}_{i\alpha}^{\dag}$, $\hat{c}_{i\alpha}$ and
$\hat{n}_{i\alpha}\equiv \hat{c}_{i\alpha}^{\dag} \hat{c}_{i\alpha}$
are respectively the creation, annihilation and number operators for
an electron with spin $\alpha$ in an orbital localized at site $i$
on a lattice of $N$ sites; the $\langle ij \rangle$ denotes pairs
$i,j$ of nearest-neighbor sites on the lattice; $U$ is the
Coulombian repulsion that operates when the two electrons occupy the
same site; and $t$ is the electron transfer integral connecting
states localized on nearest-neighbor sites. First and second terms
of Eq. (\ref{hamil}) correspond to, respectively, one-particle
$\hat{H}_{0}$ and interaction terms of Eq. (\ref{e1}).

\begin{figure*}
\psfig{figure=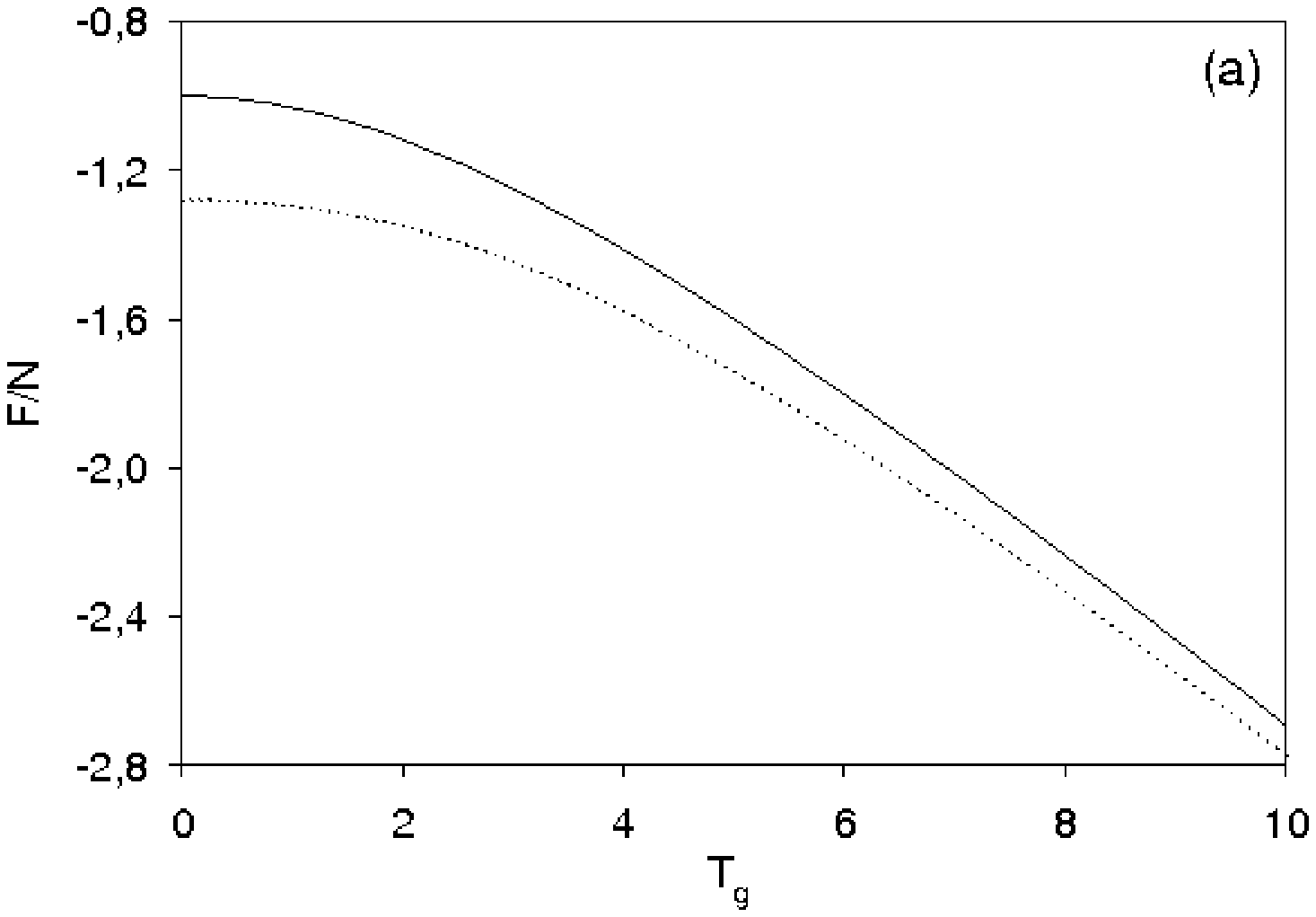,width=84mm,angle=0}
\psfig{figure=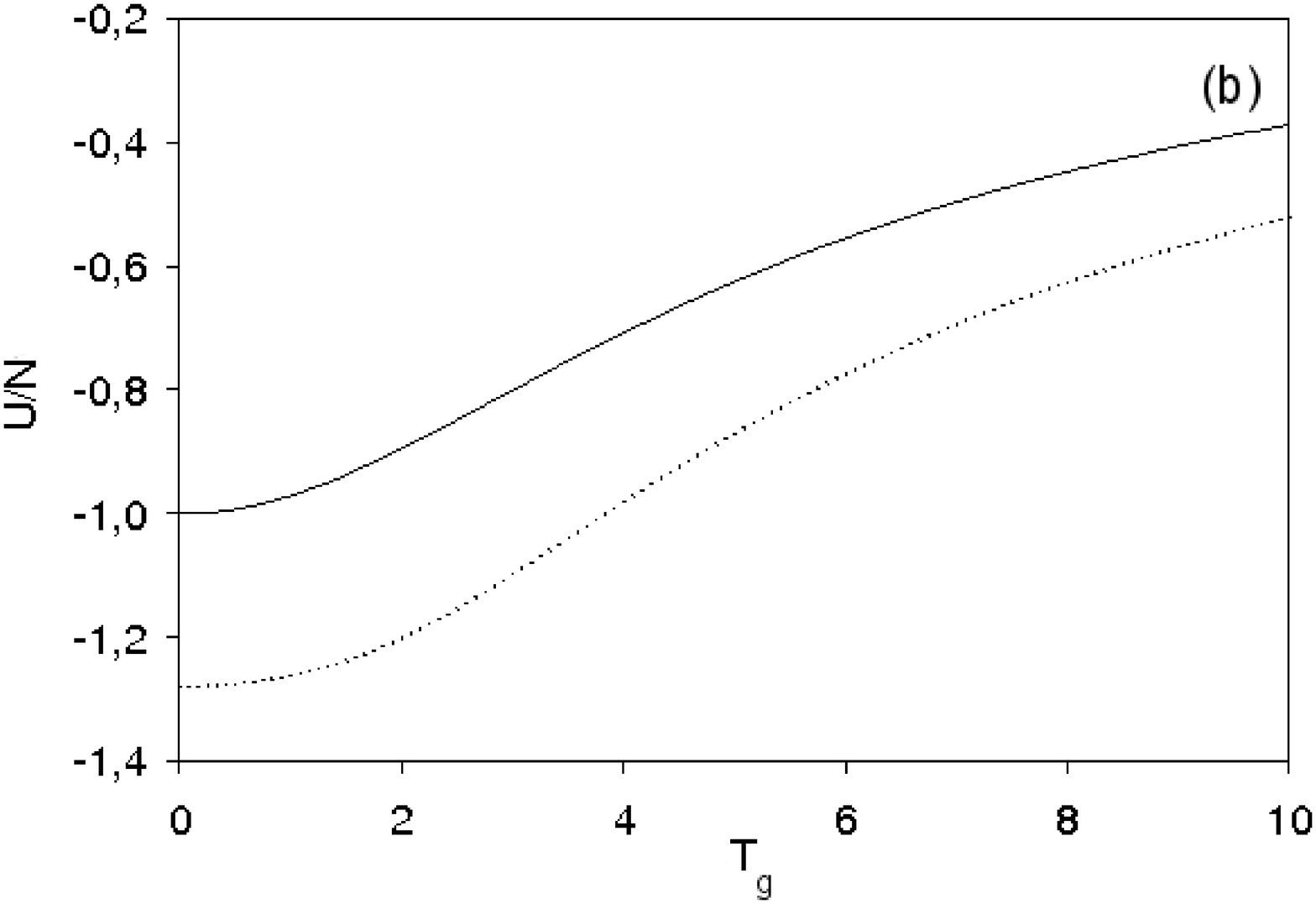,width=88mm,angle=0}
\psfig{figure=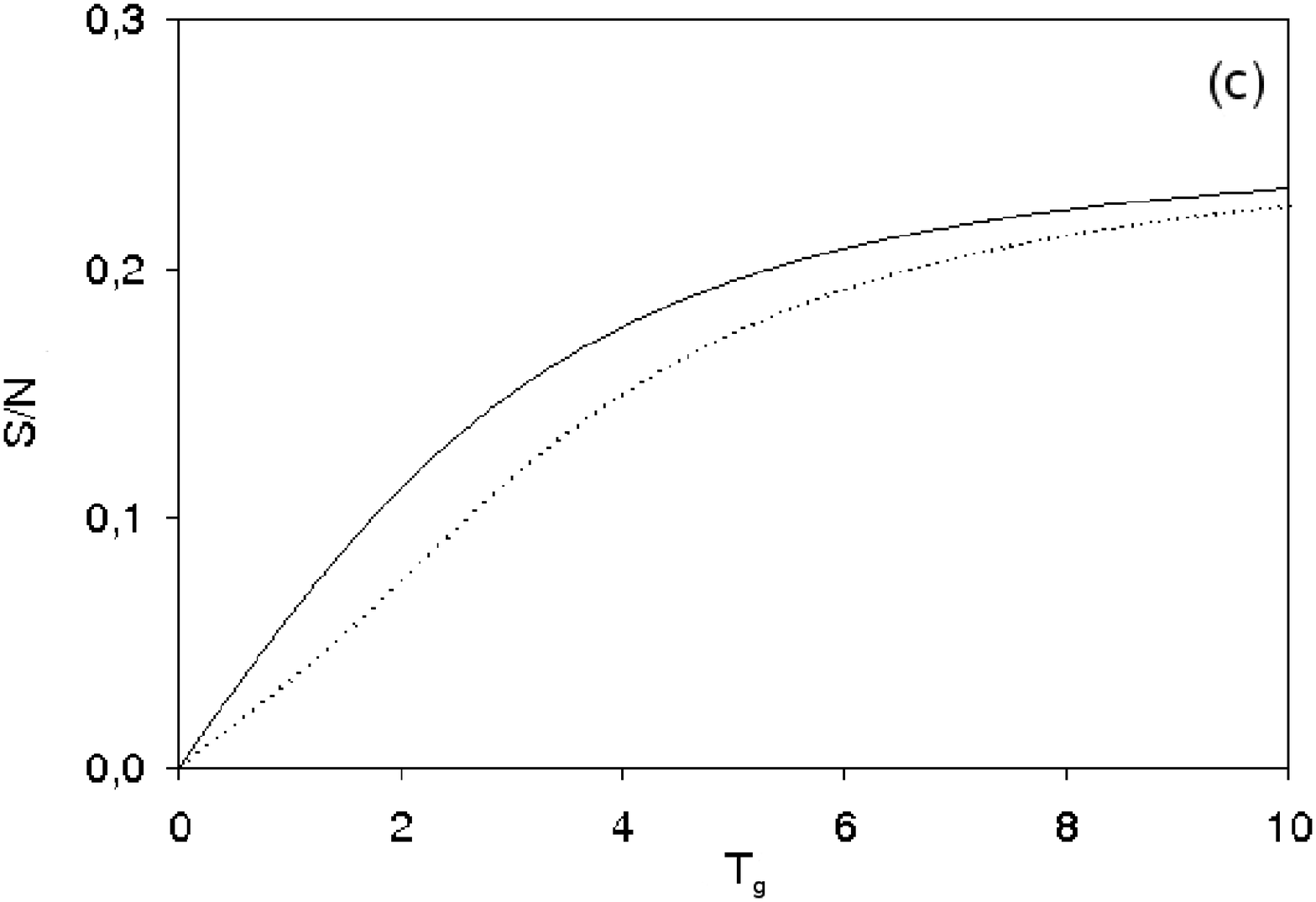,width=85mm,angle=0}
\psfig{figure=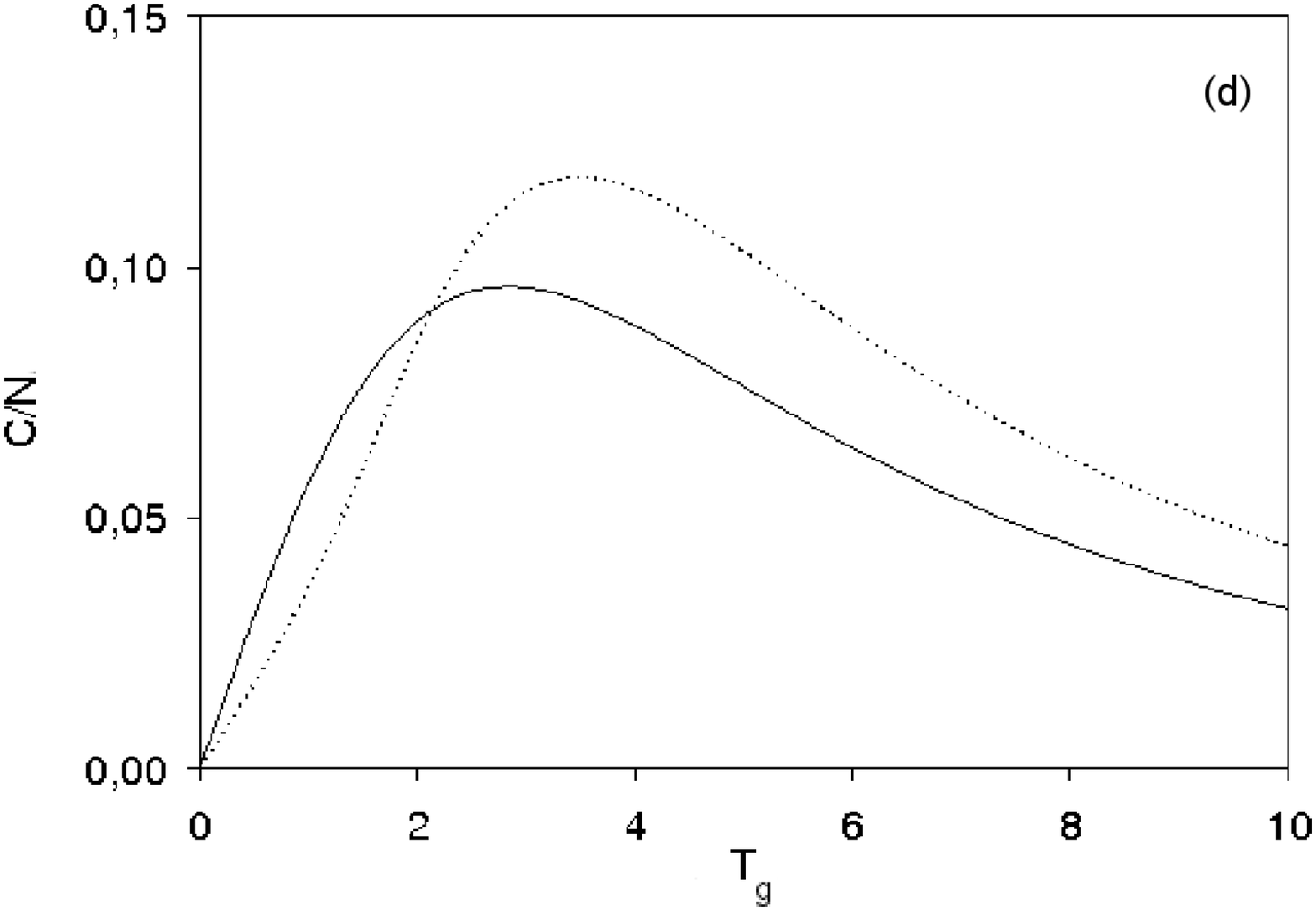,width=85mm,angle=0} \caption{Ground-state
(a) free energy $F(T_{g})$, (b) internal energy $U(T_{g})$, (c)
entropy $S(T_{g})$ and (d) heat capacity $C(T_{g})$ versus
temperature $T_g$ for the Hubbard model ($k=1$ and $t=1$). The full
line represents the case $N=2$ and two electrons, while the dotted
line represents the half-filled band for the one-dimensional case in
the thermodynamic limit ($N \rightarrow \infty$).} \label{fi2}
\end{figure*}

The problem of two electrons in two sites is the simplest example to
our approach. By using direct calculus, it is easy to obtain the
ground-state eigenvalue and eigenfunction, respectively, as
\begin{equation} \label{gs2s}
E_{0} (U) = - \frac{1}{2} (U-\sqrt{U^{2} + (4t)^{2}}),
\end{equation}
and
\begin{equation} \label{es2s}
| \psi_{0} \rangle = a_{-} | \phi_{-} \rangle + a_{+} | \phi_{+}
\rangle.
\end{equation}
where $| \phi_{\pm} \rangle $ are eigenfunctions for the case $U=0$,
with $a_{-}=\sqrt{1-a_{+}^{2}}$ and
\begin{equation} \label{as2s}
a_{+} = 2t/ \sqrt{(2\sqrt{U^{2} + (4t)^{2}} - U) \sqrt{U^{2} +
(4t)^{2}}  }.
\end{equation}
Thus, we define $T_{g}=U/kt$, and using Eqs. (\ref{ug})-(\ref{sg})
into Eqs. (\ref{gs2s}) and (\ref{es2s}), we find (from now $k=1$ and
$t=1$ for simplicity)
\begin{equation} \label{f2s}
F (T_{g}) = - \frac{1}{2} \sqrt{T_{g}^{2} + 16},
\end{equation}
\begin{equation} \label{s2s}
S (T_{g}) = \frac{T_{g}}{ 2 \sqrt{T_{g}^{2} + 16} },
\end{equation}
\begin{equation} \label{u2s}
U (T_{g}) = - \frac{8} { \sqrt{T_{g}^{2} + 16} },
\end{equation}
and
\begin{equation} \label{c2s}
C (T_{g}) = \frac{8T_{g}}{ 2 (T_{g}^{2} + 16)^{3/2} }.
\end{equation}
Figure \ref{fi2} shows curves (full lines) of $F(T_{g})$,
$U(T_{g})$, $S(T_{g})$ and $C(T_{g})$ versus the temperature $T_g$
for the expression above representing the case of two electrons on
two sites for the Hubbard model. As clearly seen in these figures,
the behavior of these new variables is exactly as expected from the
usual thermodynamics.

\begin{figure}
\psfig{figure=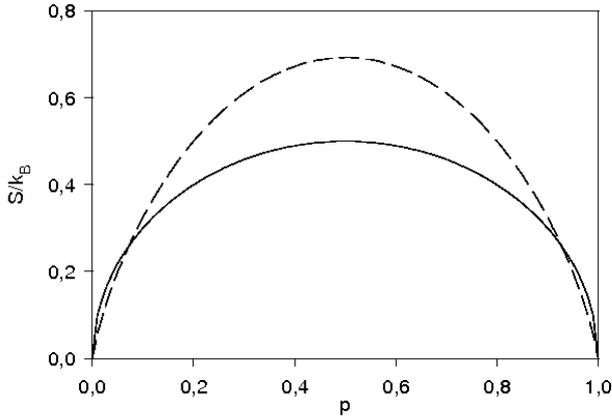,width=85mm,angle=0}

\caption{Functional forms of the entropies $S(p)$ associated with
the Boltzmann-Gibbs (dashed line) and with Eq. (\ref{sp}) for 2
states (full line).} \label{fi3}
\end{figure}

Now, let us consider the functional dependence for the entropy given
by Eq. (\ref{s2s}) in terms of the occupation probability of the
non-interacting quantum states. Using the energetic constraint (Eq.
(\ref{ug})), this dependence generates the concept of thermostat
temperature, if we focus on the canonical ensemble of the SM
formalism. It is easy to show from Eqs. (\ref{es2s})-(\ref{as2s})
that the occupation probabilities for the eigenfunctions $|
\phi_{\pm} \rangle $ of the non-interacting case are
\begin{equation} \label{p2s}
p_{\pm } (T_{g}) = \frac{1}{2} \mp \frac{2}{ \sqrt{T_{g}^{2} + 16}}.
\end{equation}
We straightforwardly obtain the entropic form
\begin{equation} \label{sp}
S(p) = \sqrt{p_{+}p_{-}},
\end{equation}
which is a concave function representing the geometric average of
the quantum states probability, where certainty corresponds to
$S=0$. Here, we can see the difference between the standard SM and
the ground-state SM. For the standard SM we always use the
Boltzmann-Gibbs entropy $S(p) = \sum_{i} p_{i} \ln{p_{i}}$, while
for the ground-state SM, this universality is broken, and the form
of the entropy depends on the particular quantum system. On the
other hand, this issue does not rule out the possibility that many
different systems fall into some basic classes exhibiting
qualitatively similar behavior. In Fig. \ref{fi3}, we show the
functional forms of the entropies associated with the
Boltzmann-Gibbs and with Eq. (\ref{sp}) assuming 2 states.

In what follows, we shall illustrate the above procedure by
addressing the exact solution for the half-filled band of the
Hubbard model for the one-dimensional case in the thermodynamic
limit. This famous solution was obtained by Lieb and Wu in the
sixties \cite{lieb} using the Bethe anzatz. Since then, this result
is considered one of the most important ones, owing to the lack of
exact solution for the Hubbard Model. The ground-state as a function
of the electron-electron interactions $U$, for $N$ sites in the
limit $N \rightarrow \infty$, is given by
\begin{equation} \label{f1d}
E_{0} (U) = - 4N \int_{0}^{\infty} \frac{J_{0}(w)J_{1}(w) dw}{w[ 1+
\exp (wU/2)] },
\end{equation}
where $J_{0}(w)$ and $J_{1}(w)$ are  Bessel functions. It is, then,
simple to obtain the quantities associated to the ground-state
thermostatistics:
\begin{equation} \label{f1d}
F (T_{g})/N = - \frac{T_{g}}{4} -4 \int_{0}^{\infty}
\frac{J_{0}(w)J_{1}(w) dw}{w[ 1+\exp (wT_{g}/2)] } ,
\end{equation}
\begin{equation} \label{s1d}
S (T_{g})/N = \frac{1}{4} - \frac{1}{2} \int_{0}^{\infty}
\frac{J_{0}(w)J_{1}(w) dw}{\cosh^{2} (wT_{g}/4)},
\end{equation}
\begin{equation} \label{u1d}
U (T_{g})/N = - \int_{0}^{\infty} \frac{J_{0}(w)J_{1}(w) f(w,T_{g})
dw}{w [ 1+\exp (wT_{g}/2)]^{2} }  ,
\end{equation}
where $f(w,T_{g})=[4+(4+wT_{g}/2)\exp (wT_{g}/2)]$ and
\begin{equation} \label{c1d}
C (T_{g})/N = \frac{T_{g}}{4} \int_{0}^{\infty} \frac{w
J_{0}(w)J_{1}(w) \sinh (wT_{g}/4) dw}{\cosh^{3} (wT_{g}/4)}.
\end{equation}
We show the behavior of $F(T_{g})$, $U(T_{g})$, $S(T_{g})$ and
$C(T_{g})$ versus the temperature $T_g$ for the solution of the
one-dimensional Hubbard model in Fig. \ref{fi2}. These curves
correspond to the dotted lines and, as well noticed from the case of
two electrons, they are also expected from the usual thermodynamics.

\section{Conclusions}
\label{concl}

In summary, we introduce an approach to solve problems of quantum
mechanics using concepts of statistical mechanics. We can consider
that taking different ground-state temperatures $T_{g}$, i.e,
different values of the interaction parameter, the particles of the
system fall into non-interacting microstates, corresponding to
different occupation probabilities for these energy levels.

We found that the functional form of the ground-state entropy
depends on the particular quantum system. The break down of
universality of the entropy is consistent with the concept of
generalized entropies \cite{tsallis} associated with a specific
quantum Hamiltonian.

Finally, the ideas presented here can eventually provide a mechanism
for new approximation methods, such as the usage of the geometric
average of the quantum states probability in the high dimensional
limit for the Hubbard model. We can envisage in further works the
study of the possibility that many different systems may fall into
some basic classes of the ground-state entropy.

\section*{Acknowledgement}

This work was supported by CNPq (Brazilian Agency).

\bibliographystyle{elsarticle-num}

\end{document}